\documentclass[useAMS,usenatbib]{mn2e}
\usepackage{graphicx}
\DeclareGraphicsExtensions{.pdf, .jpg}
\usepackage{subfigure}

\title[INTEGRAL high energy monitoring of the X-ray burster KS 1741-293]{INTEGRAL high energy monitoring of the X-ray burster KS 1741-293
\thanks{Based on observations with {\it INTEGRAL}, an ESA  project with instruments and science data centre funded by ESA member  states (especially the PI countries: Denmark, France, Germany, Italy,  Switzerland, Spain), Czech Republic and Poland, and with participation  of Russia and the USA.}}
\author[G. De Cesare, A. Bazzano, S. Mart\'inez N\'u\~nez et al.]{G. De Cesare$^{1, 2, 3}$, A. Bazzano$^{1}$, S. Mart\'inez N\'u\~nez$^{6}$, G. Stratta$^{5}$, A. Tarana$^{1, 4}$, \newauthor M. Del Santo$^{1}$ and P. Ubertini$^{1}$ \\
$^{1}$INAF-Istituto di Astrofisica Spaziale e Fisica Cosmica di Roma, via Fosso del Cavaliere 100, I-00133 Roma, Italy \\
$^{2}$Dipartimento di Astronomia, Universit\`a degli Studi di Bologna, Via Ranzani 1, I40127 Bologna, Italy \\
$^{3}$Centre d'Etude Spatiale des Rayonnements, CNRS/UPS, B.P. 4346, 31028 Toulouse Cedex 4, France \\
$^{4}$Dipartimento di Fisica, Universit\`a di Roma Tor Vergata, via della Ricerca Scientifica 1, I-00133 Roma, Italy \\
$^{5}$ASDC, via Galileo Galilei, I-00044, Frascati, Italy \\
$^{6}$GACE, Instituto de Ciencias de los Materiales, Universidad de Valencia,P.O. Box 20085, 46071 Valencia, Spain}

\begin{document}
\maketitle

\begin{abstract}
KS 1741-293, discovered in 1989 by the X-ray camera TTM in the Kvant module of the Mir space station and identified as an X-ray burster, has 
not been detected in the hard X band until the advent of the INTEGRAL observatory. Moreover this source has been recently 
object of scientific discussion, being also associated to a nearby extended radio source that in principle could
be the supernova remnant produced by the accretion induced collapse in the binary system. 
Our long term monitoring with INTEGRAL, covering the period from February 2003 to May 2005,
confirms that KS 1741-293 is transient in soft and hard X band. When the source is active,
from a simultaneous JEM-X and IBIS data analysis, we provide  a wide band spectrum from 5 to 100 keV,
that can be fit by a two component model, a  multiple blackbody for the soft emission and a Comptonized or a cut-off power law model 
for the hard component. Finally, by the detection of two X-ray bursters with JEM-X, we confirm the bursting nature of
KS 1741-293, including this source in the class of the hard tailed X-ray bursters.
\end{abstract}


\section{Introduction}
The first type I X-ray bursting sources were discovered with SAS 3 \citep{Lewin1976} and OSO-8 \citep{Swank1977}. 
 \cite{Woosley1975} and \cite{Maraschi1976} independently  discussed the origin of the phenomenon:  type 
I X-ray  bursts are explained by thermonuclear flashes of the material accreting from the companion star on the surface 
of the neutron star. All X-ray sources showing  type I bursts are low mass X-ray binaries (LMXBs).
The proprieties and the theory of the X-ray bursts are discussed in the review of \cite{Lewin1995}. 

The X-ray burster KS 1741-293 was firstly reported by \cite{Zand1991} as one of the two new transient sources near 
the Galactic Centre (GC) detected during the observations performed with the X-ray wide-field camera TTM on board the 
Kvant module of the Mir space station. KS 1741-293 was detected,  
on 3 consecutive days in the energy range 5.7$-$27.2 keV, during which it exhibited two type I X-ray bursts.  
KS 1741-293 may be identified with either  MXB 1743-29 and MXB 1742-29, two bursting sources detected in 1976 with SAS-3. 
The single peak burst profile excludes the identification of KS 1741-293 with MXB 1743-29 \citep{Zand1991}. Therefore KS 1741-293 and MXB 1742-29 
are likely to be the same source. In the BeppoSAX era (1996-2002), KS 1741-293 was detected, together with a large sample of 
galactic sources, during the Wide Field Camera (WFC) monitoring of the GC region \citep{Zand2004} at 
a peak flux of the order of 30 mCrab in the 2-28 keV energy range. 
From  the Medium Energy Concentrator Spectrometer (MECS, on board BeppoSAX) observations, \cite{Sidoli1999}
report a 2-10 keV luminosity  of the source $<10^{35}$ erg s$^{-1}$ and 10$^{36}$ erg s$^{-1}$ (corrected for absorption) 
on September 1997 and March 1998, respectively, assuming a distance of 8.5 kpc. 
KS 1741-293 was also detected by ASCA during 107 pointing  observations of a $5 \times 5$ deg$^2$ 
region around the GC showing an apparent variability by a factor of 50, while on the contrary no burst 
has been found \citep{Sakano2002}. No hard X-ray detection has been 
reported by the first gamma-ray imager SIGMA on board the GRANAT satellite and indeed the  source is not in the  
hard X-Ray SIGMA catalogue, covering the 40-100 keV range \citep{Revnivstev2004}. 
KS 1741-293 is listed in the BATSE/CGRO instrument deep sample as one of the 179 sources 
monitored along the CGRO operative life \citep{Harmon2004} even though it is not a firm detection.
KS 1741-293 is reported in the third IBIS catalogue \citep{Bird2007} at a significance level of 67 sigma with a flux of
($5.2 \pm 0.1$) mCrab in the 20-40 keV band. Recently an X-ray burst has been reported from KS 1741-293 with IBIS/ISGRI 
in the 15-25 keV band by \cite{Che2006}. It occurred on March 30, 2004.

A search for optical, infrared and radio counterparts was made by \cite{Chere1994} without finding a firm candidate. A Chandra source
inside all the KS 1741-293 high energy error circles has been proposed by \cite{Marti2007} as a possible counterpart. These authors 
discuss also a possible association with a non-thermal radio nebula that could be the supernova remnant produced
by the accretion induced collapse in the binary system. However this association is still under debate due to
the estimated age (about 500 years) of the SNR, which is very low for a LMXB.

In this work we show that, within February 2003 and May 2005, KS 1741-293 has been clearly detected in the 
hard X energy band ($> 20$ keV) with the IBIS imager during two visibility periods, while during the other observations it
appears to be in a quiescent status. Using the combined data from the X-ray monitor JEM-X and the IBIS hard X telescope, 
we obtain in the second period of visibility the wide band X-ray spectrum from about 5 keV to 100 keV. We show for the first time
the soft component simultaneously with the hard tail for this source. 
We report also the detection of two bursts with JEM-X and their temporal and spectral properties.

In section \S2 we show the observations and the data analysis tools. In section \S3 we present the data analysis results from:
flux monitoring, spectral analysis and X-ray burst analysis. Finally the conclusions are summarised in section \S4.

\section{Observations and data analysis}
\begin{table*}
  \begin{center}
    \caption{KS1741-293 IBIS observations. The visibility periods 1, 2, 3 include all the public data and 
    the core program data. Periods 4 and 5 include the core program data only. The source significance and average 
    flux are reported in the 20-40 keV energy band, the upper limits, in the same energy band, are 
    estimated at 3 sigma of significance level.}
    \vspace{1em}
    \renewcommand{\arraystretch}{1.2}
    \begin{tabular}{lrccccc}
      \hline
      Period & Rev. & Start (MJD) &  End (MJD) & Exp. (ks) & Significance & Average Flux (mCrab)\\
      \hline
      1  & 46-63     & 52698 & 52792 & 306  & 28 $\sigma$ & $9.4 \pm 0.3$ \\
      2  & 103-120 & 52871 & 52921   & 1333 & not visible & $<$ 0.5 \\
      3  & 164-185 & 53052 & 53115   & 921  & 62 $\sigma$ & $11.1 \pm 0.2$ \\
      4  & 229-249 & 53246 & 53306   & 159  & not visible & $<$ 1.4 \\
      5  & 291-307 & 53431 & 53479   & 43   & not visible & $<$ 3 \\
      \hline \\
      \end{tabular}
    \label{tab:table1}
  \end{center}
\end{table*}
\begin{table*}
  \begin{center}
    \caption{KS 1741-293 JEM-X observations. The source significance and the average flux are reported in the 
        4-15 keV (Sigma$_I$, Flux$_I$) and 15-30 keV energy bands (Sigma$_{II}$, Flux$_{II}$)}
    \vspace{1em}
    \renewcommand{\arraystretch}{1.2}
    \begin{tabular}{lrccccccc}
      \hline
      Period & Rev. & Start (MJD) &  End (MJD) & Exp. (ks) & Sigma$_I$ & Flux$_I$ (mCrab) & Sigma$_{II}$ & Flux$_{II}$ (mCrab)\\
      \hline
      1 & 46-63   & 52698 & 52792   & 66  & not visible & $<$ 1.5 & not visible & $<$ 2\\
      2 & 103-120 & 52871 & 52921   & 413 & not visible & $<$ 0.5 & not visible & $<$ 1\\
      3a & 164-169 & 53052 & 53069   & 39  & not visible & $<$ 2   & not visible & $<$ 2\\
      3b & 170-185 & 53070 & 53115   & 164 & 27 & $ 11.6 \pm 0.5 $ & 16 &  $20.2 \pm 1.5$ \\
      4 & 229-249 & 53246 & 53306   & 77  & not visible & $<$ 2   & not visible & $<$ 1.7 \\
      5 & 291-307 & 53431 & 53479   & 16  & not visible & $<$ 2.5 & not visible & $<$ 3\\
      \hline \\
      \end{tabular}
    \label{tab:table1b}
  \end{center}
\end{table*}

The X- and gamma-ray observatory INTEGRAL was launched on October 17, 2002 by the Russian PROTON launcher. The 
satellite orbits around the Earth in three days, along a highly eccentric orbit and the observing time 
is optimised by this choice. The wide-field Gamma-ray imaging and wide-band spectral capabilities of INTEGRAL 
coupled with the Core Program strategy \citep{Winkler2003},  are a powerful tool to investigate deeply 
the high energy behaviour of X-ray bursters as firstly reported by \cite{Bazzano2004}.
The scientific instruments on board are the hard X-ray and gamma-ray imager IBIS \citep{Ubertini2003} 
covering the energy band 20 keV$-$10 MeV, the gamma-ray spectrometer SPI \citep{Vedrenne2003}, that works in the same energy 
band of IBIS but is devoted to fine spectroscopy, the X-ray monitor JEM-X (3$-$35 keV) \citep{Lund2003}  and the 
optical camera OMC \citep{Mas2003}. The angular resolution of SPI is not good enough to disentangle KS 1741-293 from the nearby sources
in a crowded region as it is the Galactic Center. In the optical band the X-ray sources located in the galactic center region are generally
obscured. Thus, SPI and OMC data are not useful for our purposes and we did not analyse them. 

We have used the public data from revolution 46 (2003-02-28) to  revolution 185 (2004-04-19), and the Core 
Program data from the revolution 46 to the revolution 307 (2005-05-19).  We have selected the data from
all pointings in which KS 1741-293 is in the Fully Coded Field of View  (FCFV, equal to a square of $9 \times 9$ degrees and 
$4.8$ degrees diameter for IBIS and JEM-X respectively), where the instrument sensitivity has the best value. 

The data reduction was carried out with the release 5.1 off-line scientific analysis
(OSA) software \citep{Courvoisier2003} for IBIS and the last 6.0 release for
JEM-X. The reason of this choice is that OSA 6.0 has been released recently and
includes relevant updates in the analysis methods of JEM-X but not in IBIS.

In the OSA environment, the standard analysis pipeline requires to create firstly a set of data (observation group) and 
then to run the single tasks.  Therefore we have created an observation group for each KS 1741-293 visibility 
period and we have then performed the data reduction in two steps, obtaining firstly from the raw data the single 
pointings images and after a mosaic image combining all single images. 

During the visibility periods when the source is detectable, the IBIS mean spectrum has been obtained by the single pointing spectra.
Because the quiescent emission of the source is below the JEM-X sensitivity for each individual pointing, the JEM-X mean spectrum 
has been extracted from the mosaic image. The JEM-X bursts analysis has been performed selecting the appropriate good time intervals.

For the spectral analysis, we have used the standard fitting tool Xspec, version 11.3.1.

\section{Scientific results}
   \subsection{Flux monitoring}

\begin{figure}
\centering
\includegraphics[width=0.9\linewidth]{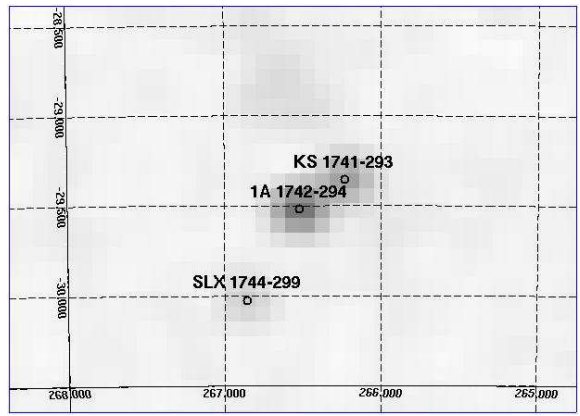}
\caption{IBIS/ISGRI mosaic, including all the data during the visibility period 3 (see Table 1),
in the 20 - 40 keV energy band. The burster KS 1741-293 has been detected during this period at 
62 sigma significance level.   
\label{fig:image_ibis}}
\end{figure}

\begin{figure}
\centering
\includegraphics[width=1.0\linewidth, height=4cm]{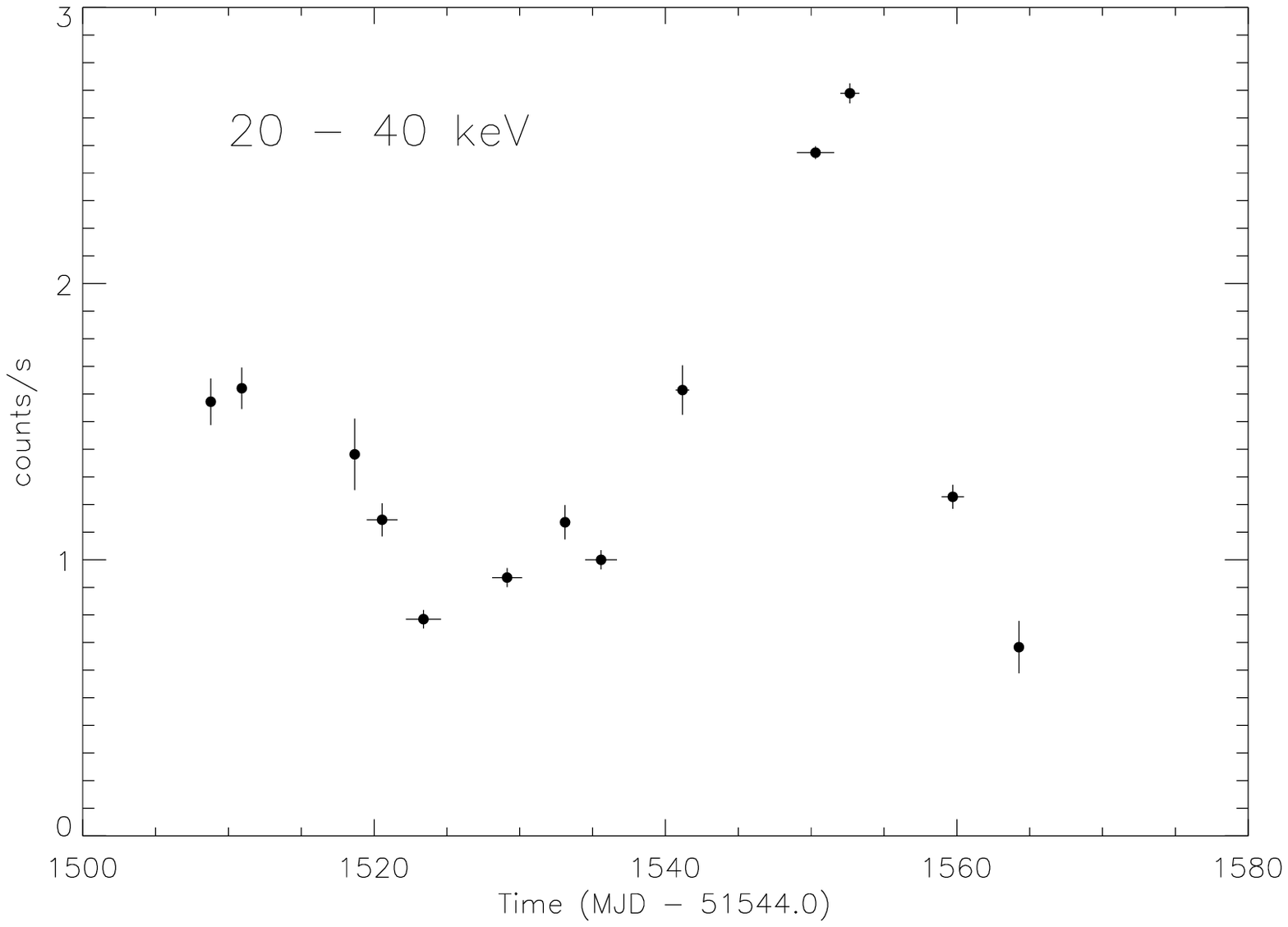}
\includegraphics[width=1.0\linewidth, height=4cm]{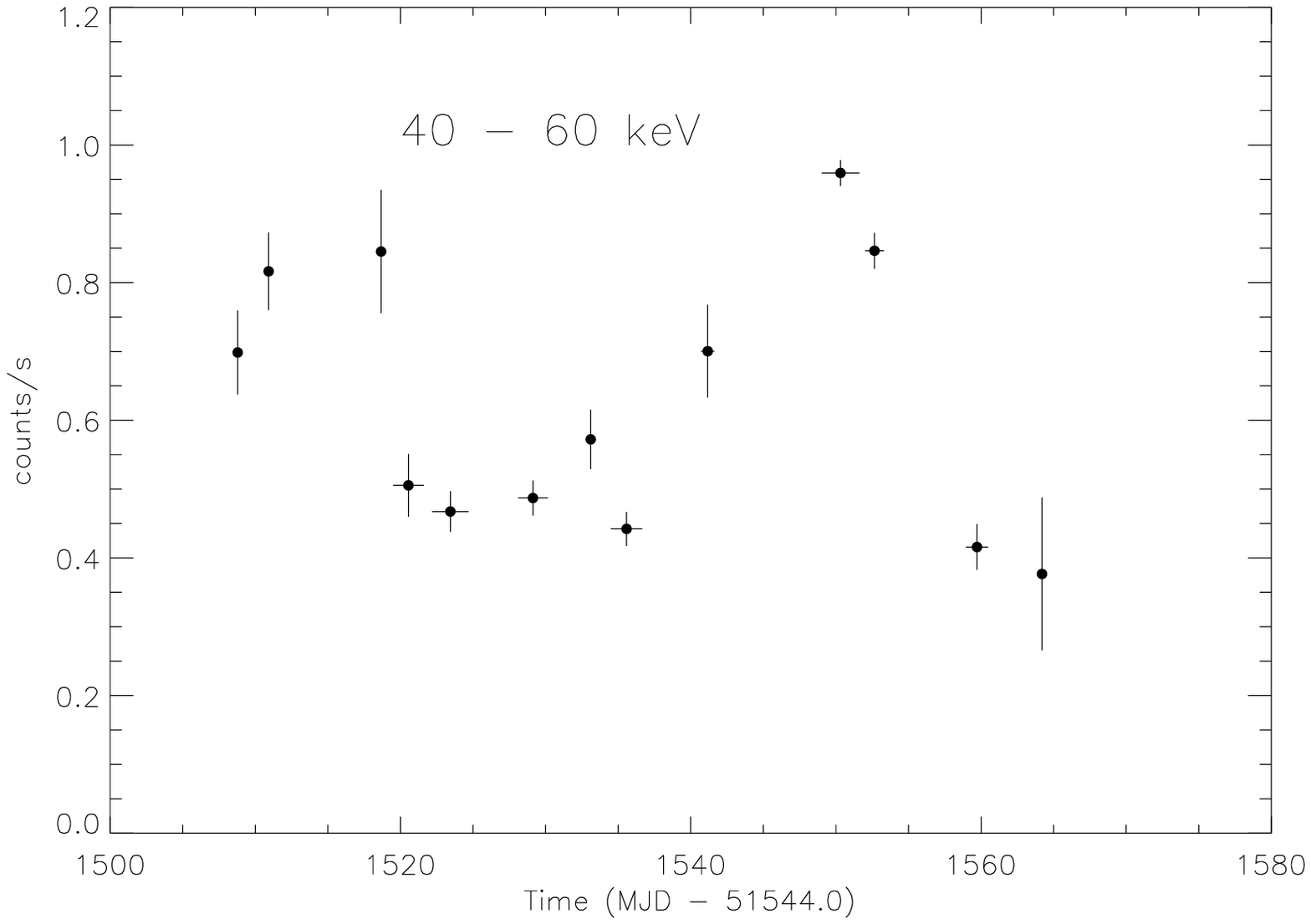}
\includegraphics[width=1.0\linewidth, height=4cm]{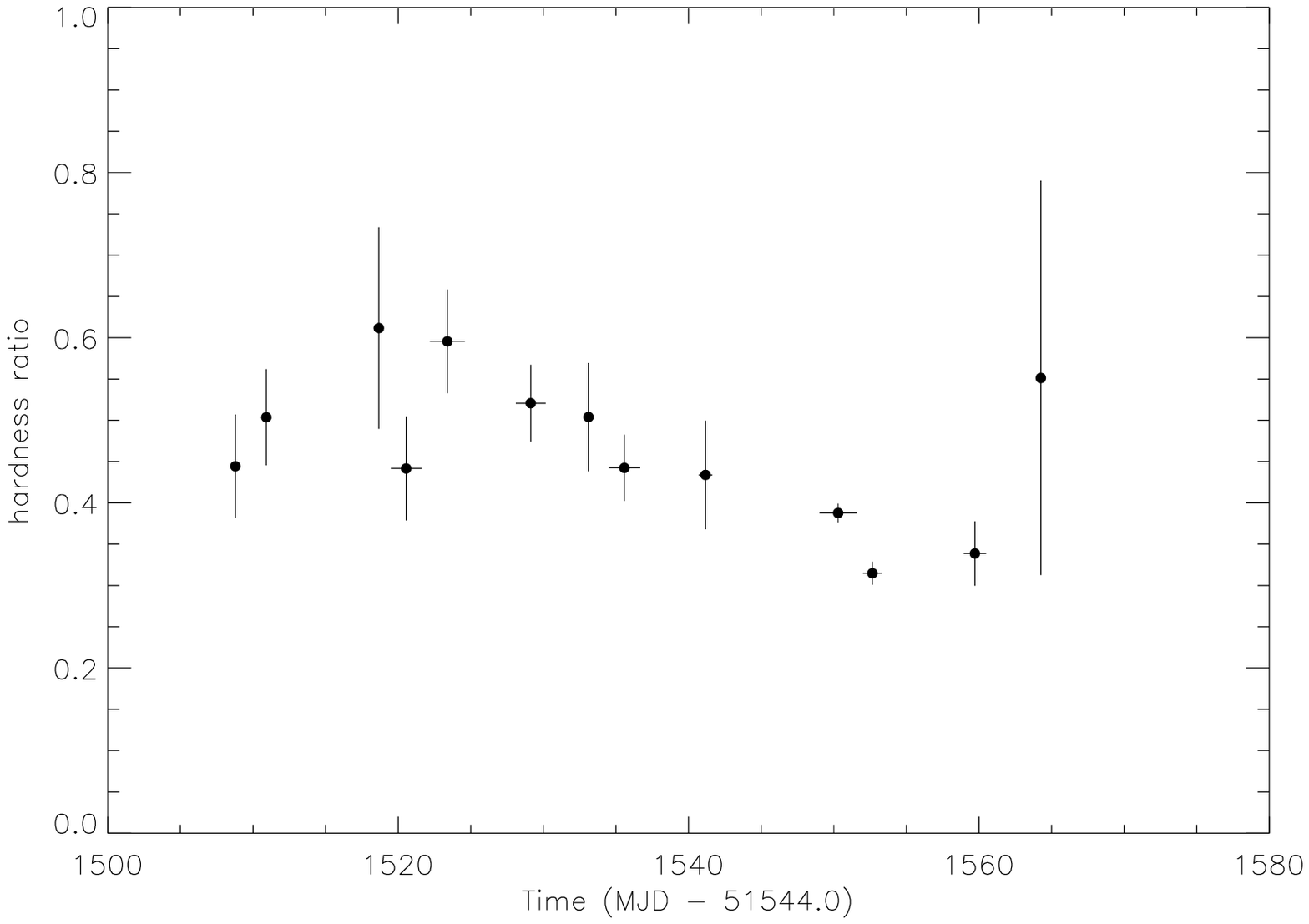}
\caption{From top to bottom: IBIS/ISGRI KS 1741-293 20-40 keV, 40-60 keV light curves and 
hardness ratio during the visibility period 3. Each point in the plots represents the average flux 
within a single INTEGRAL orbital period. The time is expressed in the 
INTEGRAL Julian Date (IJD = MJD - 51544.0). The log of the observation is quoted in table 1.} 
\label{fig:time_ibis}
\end{figure}

Table \ref{tab:table1} shows the log of our IBIS observations.
KS 1741-293 has been clearly detected in the visibility periods labelled as 1 and 3, 
with a 20-40 keV  significance of 28 sigma, and 62 sigma respectively. The average flux during these periods does 
not show any meaningful variation, ranging from  ($9.4 \pm 0.3$) mCrab during period 1 to  ($11.1 \pm 0.2$) mCrab for period 3.
For this last period the statistics are good enough to allow a study of the source's temporal behaviour.
When the source is in a quiescent status, i.e. the flux is below the instrument sensitivity, we report the 20-40 keV flux with upper limits 
estimated at 3 sigma of significance level.

The image mosaic during the period 3 is shown in figure \ref{fig:image_ibis}, 
where the sources are labelled according with \cite{Liu2001}. KS 1741-293 and SLX 1744-299 are associated with MXB 1741-29 and 
MXB 1743-29 respectively, detected by SAS-3 in 1976. 
Note that at that time 1A 1742-294 (21 arcminutes off our source) was not resolved by SAS-3.
On the contrary thanks to the good IBIS angular resolution (12 arcmin) and 
to the long exposure, we are able to obtain a good separation between these two sources. 

In figure \ref{fig:time_ibis}  we show, during period 3, the light curves in the energy band 20-40 keV, 40-60 keV and 
the hardness ratio, here defined as the ratio (count(40-60 keV)/count(20-40 keV)) between the high energy to low energy band.  
To increase the statistics of the single IBIS pointings, we 
have grouped all the IBIS pointing data associated to one satellite revolution. 
The bin width used for the flux integration is different from one revolution to another, ranging from about 30 min (one IBIS pointing) to 30 hours.
This explains the differences in the error bars. The light curve in the 20-40 keV shows a variability of about a factor of four, smoothly 
increasing from 0.8 counts/s to 2.7 counts/s in 30 days. Moreover we find an indication of spectral softening at IJD equal to about 1550.

The JEM-X observations are reported in the table \ref{tab:table1b}. JEM-X was operated with JMX2 until revolution 170 and since then 
it is operating mainly with JMX1. Our source is detected in the 4-15 keV and 15-30 keV energy bands at a significance level of
27 and 16 respectively in period 3 from March 6 (revolution 170, IJD= 1526) to April 20 2004 (revolution 185, IJD = 1571),
while is not detected during periods 1 and 3a (table 2). The lack of JEM-X detection
during these periods can be explained by both the less exposure and by the source flux variability
on temporal scales of the order of few days as we observe during period 3.
The JEM-X exposures during periods 1 and 3a are a factor of 2.5 and 4.2 lower than in period 3b.
Assuming a constant flux during all the observations and taking into account the less exposure,
we would expect a signal at 17 and 13 sigma in the period 1 and 3a, respectively. However, the
flux is not constant. In fact, the 20-40 keV IBIS light curve during the period 3 (figure 2, top plot)
shows that flux can varies more than a factor of two, reaching its maximum (26 mCrab) during
period 3b, that is when we have the JEM-X detection.
The mean 20-40 keV flux in period 3b (18 mCrab) is about two times the one during
period 1 (9.4 mCrab). Thus, we would expect from JEM-X in period 1 a signal that is a factor of
two less than the above calculations (17 sigma), that is of about 9 sigma.
This is a marginal significance level for a detection with a coded mask telescope. The same
arguments can be applied to the lack of detection in period 3a. Moreover,
during period 1 and 3a the signal could be even less significant since we have indication of
spectral softening during period 3b (figure 2, bottom plot) when the source is detected by JEM-X.

   \subsection{Spectral analysis}
\begin{figure}
\centering
\includegraphics[width=6cm,height=8cm,angle=-90]{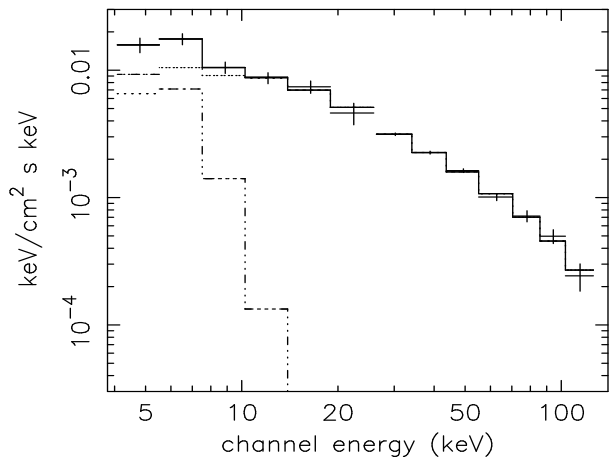}
\caption{IBIS/ISGRI and JEM-X KS 1741-293 average spectrum obtained in the 3th visibility period.The solid line is the
best fit model assuming a multi blackbody plus a Comptonized model.}
\label{fig:spectrum_ibis}
\end{figure}

\begin{table*}
  \begin{center}
    \caption{Best fit parameters from a simultaneous IBIS JEM-X spectral analysis during period 3 assuming a multicolor disk blackbody 
component plus a cut-off power law (cutoffpl in Xspec) or a Comptonized model (CompTT in Xspec)
with a plan geometry. $T_{in}$ is the temperature at 
inner disk radius, $E_{cut}$ is the cut-off energy, $T_0$ is the input soft photon (Wien) temperature, $kT_e$ and $\tau$ are the plasma 
temperature and optical depth, respectively. The errors are at 68 \% significance level.
}
    \vspace{1em}
    \renewcommand{\arraystretch}{1.2}
    \begin{tabular}[h]{cccccccccc}
      \hline
      \hline
Model & $N_H$& $T_{in}$ & $\Gamma$ & $E_{cut}$ & $T_0$  & $kT_e$ & $\tau$ & Flux (2-100 keV) & $\chi^2_{\nu}(\nu)$ \\
      &  $10^{22}$cm$^{-2}$     &  keV     &          &   keV     &  keV   &   keV  &       & $10^{-10}$erg cm$^{-2}$ $s^{-1}$ & \\
      \hline
const*wabs*(diskbb+cutoffpl) & $35\pm10$ & $1.0\pm0.2$  & $2.0\pm0.2$ & $90^{+40}_{-20}$ & - & - & - & 7.7 & 1.1(6) \\
      \hline
const*wabs*(diskbb+compTT) & $31^{+12}_{-8}$ & $1.1\pm0.2$ & - & - & 0.1 & $28^{+16}_{-4}$ & $1.0\pm0.5$ & 7.7 & 1.0(6) \\
	\hline
      \end{tabular}
    \label{tab:table2}
  \end{center}
\end{table*}

The IBIS/ISGRI average spectrum has been extracted in the 20-150 keV energy range during the two periods corresponding to source detection, namely 
period 1 and period 3 (see table \ref{tab:table1}). We fitted the extracted spectra with a simple power law, a cut-off power law and a Comptonized 
model with Xspec  (v.11.3.2). During Period 1 we find that all these models provide a good agreement with the data, with a marginal evidence of 
improvement by using the cut-off power law or a Comptonized model (the probability of a chance improvement in $\chi^2$ 
is $\sim7\%$ using the F-test). 
During period 3, the increased statistics due to the longer exposure than for period 1, enable us to confidently exclude a simple power law 
model (with reduced $\chi^2_{\nu}=3.3$ with 5 degrees of freedom). We obtain a comparable good fit good fit with a cut-off power law ($\chi^2_{\nu}(\nu)=0.5(4)$) and with a 
Comptonized model ($\chi^2_{\nu}(\nu)=0.5(4)$).
We find no evidence of spectral variability between Period 1 and Period 3 in the 20-150 keV energy range. 

In order to constrain the model parameters, we extended our spectral analysis to low energies of 4 keV thanks to the simultaneous 
detection with JEM-X during Period 3. Analysis at low energies of this source have already been performed by \cite{Sidoli1999} 
with BeppoSAX/MECS observations taken on the 31st of March 1998. From their analysis, the 2-10 keV spectrum was equally well 
modelled by a simple power law, a thermal bremmstrahlung and a black body. 
Interestingly, in the two former cases, high  photoelectric absorption was required by the data, with an equivalent hydrogen column density $N_H$ of 
a few $10^{23}$cm$^{-2}$. 
 
We then performed a simultaneous analysis of the JEM-X and IBIS/ISGRI spectra on the basis of the previous results. A constant normalisation factor 
has been introduced in the fitting models in order to take into account the systematic errors in the knowledge of the absolute JEM-X and IBIS/ISGRI inter-calibration. 
We find that, assuming a cut-off power-law model at high energies, none of the previous models used to fit the low energies (Sidoli et al. 1999) can 
fit the JEM-X data. On the contrary, a multicolor disk blackbody component (\rm diskbb in \rm xspec), typically invoked to model the emission from an accretion 
disk, provides a good fit to the data. The available statistics do not allow us to disentangle among a cut-off power-law or a Comptonized model at high 
energies, thus we obtained similar results also assuming a Comptonized model. At the same time, we were able to confidently exclude a simple power 
law model (with $\chi^2(\nu)=4.8(7)$) rather than a cut-off power-law or a Comptonized model. The best fit model parameters from these analysis are summarised in Table 2. 

We also compared the 2-10 keV flux obtained on March 1998 from the BeppoSAX/MECS observations (Sidoli et al. 1999) with our estimate from the best fit models 
in the same energy range. We find a 2-10 keV flux of $2.4\times10^{-10}$ cm$^{-2}$ s$^{-1}$ that is consistent with the previous measure.

Assuming the distance of 8.5 Kpc for this source as quoted by  \cite{Sidoli1999}, we computed the source luminosity from the flux measured during the period 3 in 
the 20 - 100 keV energy band as 1.7 $\times$ $10^{36}$ erg s$^{-1}$.

   \subsection{X-ray bursts}

\begin{figure}
\centering
\includegraphics[width=0.9\linewidth]{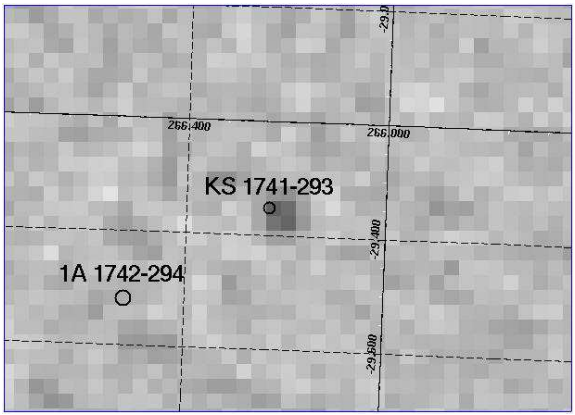}
\caption{JEM-X KS 1741-293 image in the 4 - 15 keV energy band during the bursts occurred in the revolution 53 (March 22 2003) with superimposed
the IBIS positions with a 0.6 arcmin 90 \% error circle radius. The JEM-X error circle radius with the same confidence level 
(not plotted in this picture) is equal to 3 arcmin. The 1A 1742-294 source is not detected due to a very short (about 20 seconds) integration time.}   
\label{fig:image_jemx}
\end{figure}

\begin{figure}
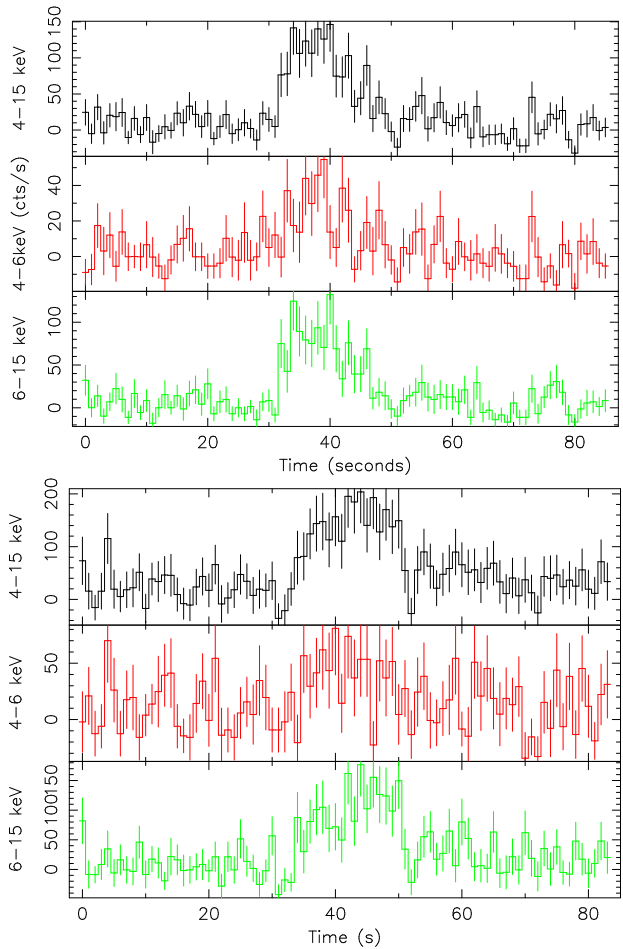

\centering
\includegraphics[width=0.74\linewidth, angle=-90]{fig5a.ps}
\includegraphics[width=0.74\linewidth, angle=-90]{fig5b.ps} 
\caption{Light curves of the two bursts detected during the revolution 53 (upper plot) and 63 (bottom plot) in the 4-15 keV band. The zero time corresponds
to 1176.359 and 1207.412 IJD for the revolution 53 and 63 respectively.}   
\label{fig:time_jemx}
\end{figure}

\begin{figure}
\centering
\includegraphics[height=6cm]{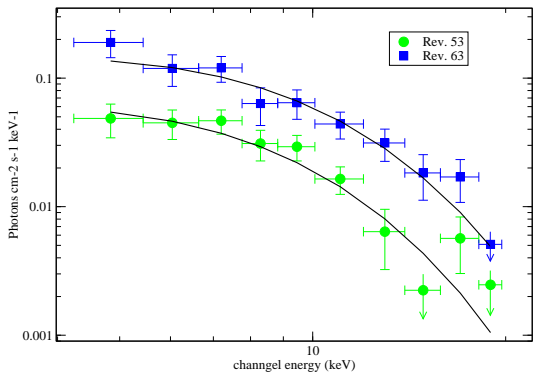}
\caption{JEM-X KS 1741-293 spectrum obtained during the bursts detected in the revolutions 53 and 63. The solid lines are the best fit
assuming a blackbody model.}
\label{fig:spectrum_jemx}
\end{figure}

Two X-ray bursts from KS 1741 has been detected in the 4-15 keV band by the JEM-X telescope during the
integral revolution 53 (March 22 2003) and 63 (April 22 2003). On the instrument  image of the first burst
(see fig. \ref{fig:image_jemx}) we have superimposed the IBIS position as it is reported with good accuracy
(0.6 arcmin 90\% error circle) in the survey of \cite{Bird2007}. Note that the JEM-X position accuracy (3 arcmin 90\% error circle)
is lower then the IBIS ones due to shorter exposure time. As the JEM-X position of KS 1741-293, obtained during the X-ray burst,
is in agreement with IBIS, the burst is firmly associated to our source.

The one-second resolved light curves obtained from JEM-X  data  exhibit two X-ray bursts in the energy range 4 - 15 keV,
one in revolution 53 (science window 58) and the other in revolution 63 (science window 92).
 Both bursts lasted around 20 s, with a maximum flux of 1.3 and 2.1 Crab respectively. 
The burst morphology  (fig. \ref{fig:time_jemx}) confirms previous observations with a single peaked time profile unlike
double peaked time profile reported for MXB 1743-29 \citep{Zand1991}.

The JEM-X spectra during the bursts have been extracted selecting the Good Time Intervals (GTI) 1176.359-1176.360 and 1207.4116-1297.4120 IJD 
for the $1^{st}$ and $2^{nd}$ bursts respectively. Both spectra have been fitted with a black body model (Fig. 6), with  
a temperature kT$_{bb}$ equal to $2.1^{+0.3}_{-0.3}$  keV for the first burst and $2.5^{+0.5}_{-0.4}$ keV 
for the second one  at 90 \% of confidence level. The 4-20 keV fluxes are 
3.6 $\times$ $10^{-9}$ erg cm$^{-2}$ s$^{-1}$ and 1.1 $\times$ $10^{-8}$  erg cm$^{-2}$ s$^{-1}$, respectively. Assuming a distance of 8.5 kpc, the radius 
of the first and the second bursts are $3.9^{+1.1}_{-1.0}$ km and $4.9^{+2.1}_{-1.5}$ km respectively. 

The burst emission is not detected by IBIS. This  lack of high  energy ($>$ 20 keV) detection can be 
explained by the soft spectrum of the source emission during the burst activity and by the short exposure. 
Indeed, taking into account the IBIS sensitivity, we estimate a 2 sigma flux upper limit of 140 mCrab in 
the 20-40 keV energy band for a 20 seconds exposure (i.e. the burst time interval). Extrapolating the JEM-X 
burst spectrum we  obtain in this band a flux of 9 mCrab and 7 mCrab for the first and the second burst 
respectively. These values are significantly below the IBIS 20 seconds upper limit.

\section{Conclusions}
Despite KS 1741-293 being reported for the first time many years ago \citep{Zand1991}, this source was
poorly studied, in particular at high energies.
This is mainly due to the source faintness, both in the soft and hard X-ray energy range, 
and to its position in the Galactic Centre crowed region, requiring good angular resolution. 
The IBIS angular resolution, good sensitivity, large field of view and long exposure in the region of the 
Galactic Center are then appropriate to fulfil this task. In particular the IBIS angular resolution ($12'$) allowed us to clearly resolve 
KS 1741-293 from other X-ray sources in the field of view, especially from the nearest source 1A 1742-294. 

Using Open Time and Core Programme data, we have monitored, with a mCrab sensitivity in the 20-40 keV band, the hard X-ray emission
for a time period of more then two years. We have obtained a clear IBIS detection only during the visibility periods 
52698-52792 MJD and 53052-53115 MJD, showing that hard X-ray emission from this source is not persistent. 

The measured orbital periods for LMXBs range from a fraction of hour to tens of hours (see \cite{White1995}).
During the $3^{rd}$ visibility period the source shows a smooth flux variation by a factor of four, without evidence 
of periodicity in the light curve on a time scale of 60 days. During the first visibility period there is no evidence 
of source flux variation.

We have obtained for the first time a wide band (from 5 to 100 keV) spectrum using the simultaneous JEM-X and IBIS
data. The spectrum is fitted with a two component model. While the blackbody soft component could originate from the surface 
of an accretion disk, the neutron star surface, or both, the hard tail, fitted by a cut-off power law, or  by a Comptonized 
model \citep{Titarchuk1994}, is due to a Comptonization of soft photons in a hot plasma around the neutron star.
 
We have detected two type I X-ray burst with JEM-X, at a position clearly consistent with the IBIS KS 1741-293 detection.
The temporal analysis with JEM-X confirms the single peak bursts of KS 1741-293, as firstly reported by \cite{Zand1991}.  

The spectral proprieties of the hard tailed LMXBs are discussed in  \cite{Disalvo2002}. KS 1741-293 can be included in this
class of sources.

\section*{Acknowledgements}
Authors are grateful to the anonymous referee for the useful comments.
Authors thanks M. Federici (IASF/Rome) for the continuous effort to update the INTEGRAL archive in Rome.
This work has been supported by Italian Space Agency by the grant I/R/046/04.

\end{document}